\documentclass[aps,prl,showpacs,twocolumn,groupedaddress]{revtex4}

\usepackage{graphicx}% Include figure files
\usepackage{dcolumn}% Align table columns on decimal point
\usepackage{bm}% bold math

\begin{document}

\preprint{APS/123-QED}

\title{Magnetic Phase Diagram of the Breathing Pyrochlore Antiferromagnet LiGa$_{1-x}$In$_{x}$Cr$_4$O$_8$}% Force line breaks with \\

%\author{Ann  Author}
% \altaffiliation[Also at ]{Physics Department, XYZ University.}%Lines break aut%omatically or can be forced with \\
%\author{Second Author}%
% \email{Second.Author@institution.edu}
%\affiliation{%
%Authors' institution and/or address\\
%This line break forced with \textbackslash\textbackslash
%}%
%
%\author{Charlie Author}
% \homepage{http://www.Second.institution.edu/~Charlie.Author}
%\affiliation{
%Second institution and/or address\\
%This line break forced% with \\
%}%

\author{Yoshihiko Okamoto$^{1,*,\dagger}$, G\o ran J. Nilsen$^{2}$, Taishi Nakazono$^{1}$, and Zenji Hiroi$^{1}$}
\affiliation{
$^{1}$
Institute for Solid State Physics, University of Tokyo, Kashiwa 277-8581, Japan\\
$^{2}$Institut Laue-Langevin, 6 rue Jules Horowitz, Bo\^{i}te Postale 156, 38042 Grenoble Cedex 9, France\\
%$^{2}$CREST, Japan Science and Technology Agency (JST), Japan\\
%$^{2}$Department of Advanced Materials, University of Tokyo and CREST-JST, 5-1-5 Kashiwanoha, Kashiwa, Chiba 277-8561, Japan\\
%$^{4}$Institute for Solid State Physics, University of Tokyo, 5-1-5 Kashiwanoha, Kashiwa, Chiba 277-8581, Japan
}

\date{\today}% It is always \today, today,
             %  but any date may be explicitly specified

\begin{abstract}
The spinel oxides LiGaCr$_4$O$_8$ and LiInCr$_4$O$_8$ contain size-alternating pyrochlore lattices of spin-3/2 Cr$^{3+}$ tetrahedra with different magnitudes of alternation. We show here that the solid solutions LiGa$_{1-x}$In$_{x}$Cr$_4$O$_8$ between these two ``breathing'' pyrochlore compounds display (i) rapid suppression of magnetic and structural transitions upon doping the end members, (ii) spin-glass-like freezing above 2 K in the range 0.1 $\lesssim x \lesssim$ 0.6, and (iii) apparent spin-gap behavior for $x \gtrsim$ 0.7. Furthermore, no transitions are observed above 2 K at $x \sim$ 0.9, where magnetic susceptibility remains finite at 2 K and magnetic heat capacity shows a quadratic temperature dependence at 1$-$5 K. Our work shows that breathing pyrochlore compounds provide a unique opportunity for studying both geometrical frustration and bond alternation. 
\end{abstract}

\pacs{Valid PACS appear here}% PACS, the Physics and Astronomy
                             % Classification Scheme.
%\keywords{Suggested keywords}%Use showkeys class option if keyword
                              %display desired
\maketitle

Geometrical frustration and bond alternation have been important topics in the field of magnetism over the last decades. The former is expected when antiferromagnetically interacting spins are arranged on lattices with triangular motifs. Since all three spins in a triangle cannot be antiparallel with each other, one finds conventional magnetic order replaced by more exotic states. One of the central issues in geometrical frustration is to observe a quantum spin liquid such as the resonating valence bond state, where spin pairs are strongly entangled with each other, preserving both translation and spin rotation symmetries~\cite{Balents}. Many frustrated antiferromagnets have so far been investigated as candidates, including several copper minerals and a vanadium oxyfluoride for the kagome lattice~\cite{herb,vesi,V-kagome}, two organic triangular lattice systems~\cite{ET, dmit}, and a hyperkagome iridium oxide~\cite{Na4Ir3O8}. 

Bond alternation provides us with another ingredient to reach exotic states of matter. It has mainly been studied in one-dimensional antiferromagnetic systems, where bonds with strong and weak antiferromagnetic interactions, $J$ and $J^{\prime}$, alternate along the chain. The spin-1/2 chain shows a gapless Tomonaga-Luttinger liquid state when it is uniform, while a spin-gapped state with a spin singlet pair on the $J$ bond results when a finite alternation is introduced ($J > J^{\prime}$). On the other hand, in the case of spin-1 chains, a gapped Haldane state appears for zero or small alternations, while a dimer singlet state (also gapped) is stabilized at large alternations. Since the physical origins of these spin gaps are essentially different, there exists a gapless state at a quantum critical point $J^{\prime}$/$J$ = 0.6 between the two phases~\cite{Affleck, BA-1}. Many Ni complexes have been investigated as model compounds for the alternating spin-1 chain. For example, [\{Ni(333-tet)($\mu$-N$_3$)\}$_n$](ClO$_4$)$_n$ shows a gapless magnetic susceptibility probably related to the critical point~\cite{Ni}. 
The critical ratio between the analogous gapped states for $S$ = 3/2 seems to be reduced to $\sim$0.4~\cite{BA-2}, for which Cr$_2$BP$_3$O$_{12}$ has been studied as a model compound~\cite{CrBPO}. 

We have recently noted that the two ingredients above, geometrical frustration and bond alternation, coexist in the three-dimensional breathing pyrochlore (BP) lattice, in which small and large tetrahedra alternate in a corner-sharing geometry, as shown in the inset of Fig. 2(a). The BP lattice is realized in LiGaCr$_4$O$_8$, LiInCr$_4$O$_8$~\cite{BP-2}, and Ba$_3$Yb$_2$Zn$_5$O$_{11}$~\cite{BYZO}. The former two compounds crystallize in the A-site ordered spinel structure with the space group $F\bar{4}3m$~\cite{BP-3}, a subgroup of $Fd\bar{3}m$ for conventional spinel oxides. This symmetry lowering is due to the zinc-blende-type order of the Li$^+$ and Ga$^{3+}$/In$^{3+}$ ions on a diamond lattice, which causes chemical pressure and leads to bond alternation in the Cr$^{3+}$ pyrochlore lattice. There are large antiferromagnetic couplings between Cr spins, as evidenced by negative Weiss temperatures of $\theta_{\mathrm{W}}$ = $-$659 and $-$332 K for the Ga and In compounds, respectively. Ba$_3$Yb$_2$Zn$_5$O$_{11}$ also crystallizes in a cubic structure with the space group $F\bar{4}3m$, where pseudospin-1/2 Yb$^{3+}$ ions form a BP lattice.

The spin Hamiltonian of the BP lattice is written as $H$ = $J$$\Sigma_{ij}\mathbf{S}_{i} \cdot \mathbf{S}_j$ $+$ $J^{\prime}$$\Sigma_{ij}\mathbf{S}_{i} \cdot \mathbf{S}_j$, where the summations over $ij$ in the first and second terms run over the Cr$-$Cr bonds of small and large tetrahedra with nearest-neighbor magnetic interactions $J$ and $J^{\prime}$, respectively. It is intuitively expected that the magnetic properties of the BP antiferromagnets strongly depend on the magnitude of size alternation. Hence, we define the ratio between $J$ and $J^{\prime}$ as the breathing factor $B_{\mathrm{f}}$ = $J^{\prime}$/$J$. $B_{\mathrm{f}}$ = 1 yields a uniform pyrochlore lattice, while $B_{\mathrm{f}}$ = 0 results in an array of isolated tetrahedra. The ground state of a tetrahedron is a tetramer singlet, a quantum superposition of two singlet pairs on a tetrahedron~\cite{tetramer}. The $B_{\mathrm{f}}$'s of the Ga and In compounds are estimated to be $\sim$0.6 and $\sim$0.1 using an empirical relationship between the strength of magnetic interactions and the Cr$-$Cr distance~\cite{BP-2}. Thus, the Ga compound is almost halfway between the uniform pyrochlore and the isolated tetrahedral limits, while the In compound lies close to the latter. 

The magnetic properties of LiGaCr$_4$O$_8$ and LiInCr$_4$O$_8$ are significantly different, reflecting the difference in $B_{\mathrm{f}}$~\cite{BP-2}. The former shows antiferromagnetic short-range order below $\sim$45 K in magnetic susceptibility, much like conventional Cr spinel oxides, while the In compound shows spin-gap behavior below $\sim$65 K. At low temperatures, both compounds exhibit definite phase transitions evidenced by sharp peaks in heat capacity~\cite{BP-2}. This is also the case for the conventional spinel oxide ZnCr$_2$O$_4$, where long-range magnetic order accompanied by lattice distortion is induced by strong spin-lattice coupling~\cite{Cr-1}. 

\begin{figure}
\includegraphics[width=8.7cm]{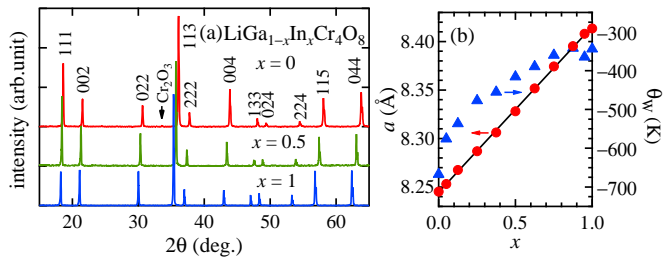}% Here is how to import EPS art
\caption{\label{fig1} 
(color online) (a) Powder XRD patterns of LiGa$_{1-x}$In$_x$Cr$_4$O$_8$ polycrystalline samples with $x$ = 0 (top), 0.5 (middle), and 1 (bottom). Cu K$\alpha_1$ and K$\alpha_2$ radiations were used. (b) $x$ dependences of lattice constant $a$ (left) and Weiss temperature $\theta_{\mathrm{W}}$ (right) for LiGa$_{1-x}$In$_x$Cr$_4$O$_8$ polycrystalline samples. The solid line shows a linear fit, giving $a$/\r{A} = 8.2446(8) + 0.171(13)$x$. 
}
\end{figure}

We prepared seventeen polycrystalline samples of LiGa$_{1-x}$In$_x$Cr$_4$O$_8$ ($x$ = 0, 0.05, 0.075, 0.1, 0.125, 0.25, 0.375, 0.5, 0.625, 0.75, 0.875, 0.9, 0.925, 0.94, 0.95, 0.975, and 1) by the solid-state reaction method. LiGaCr$_4$O$_8$ and LiInCr$_4$O$_8$ powders were mixed in a $1 - x$ : $x$ molar ratio and sintered at 1100 $^{\circ}$C for three days with intermediate grindings. Sample characterization was performed at room temperature using a RINT-2000 (Rigaku) powder X-ray diffractometer with Cu K$\alpha$ radiation. All diffraction peaks observed in the powder XRD pattern of every sample are as sharp as those of the end members and can be indexed on the basis of a cubic structure of the space group $F\bar{4}3m$, as typically shown for $x$ = 0, 0.5, and 1 in Fig. 1(a). The lattice constant varies almost linearly with $x$ [Fig. 1(b)], indicating that complete solid solutions have been obtained. 

Magnetic susceptibility was measured in a Magnetic Property Measurement System. Heat capacity was measured by the relaxation method in a Physical Property Measurement System (both Quantum Design). Polarized diffuse neutron scattering was performed on the $^7$Li-enriched $x$ = 0.9 and 1 samples on the D7 spectrometer at the Institut Laue-Langevin (ILL) using $\lambda$ = 4.84 \r{A} neutrons. The magnetic component of the total cross section was extracted by $xyz$ polarization analysis~\cite{PolarizedNeut}. 

Curie-Weiss fits of the magnetic susceptibilities of polycrystalline samples of LiGa$_{1-x}$In$_x$Cr$_4$O$_8$ to the equation $\chi$ = $N_{\mathrm{A}}$$g^2$$\mu_{\mathrm{B}}^2$$S(S + 1$)/3$k_{\mathrm{B}}$($T - \theta_{\mathrm{W}}$), where $N_{\mathrm{A}}$, $g$, $\mu_{\mathrm{B}}$, and $k_{\mathrm{B}}$ are the Avogadro constant, the Lande g factor, the Bohr magneton, and the Boltzmann constant, respectively, have been carried out in the temperature range of 300$-$350 K. This yields $g$ = 1.90$-$2.09 for $S$ = 3/2 with systematically varying $\theta_{\mathrm{W}}$ between $\theta_{\mathrm{W}}$ = $-$667 and $-$344 K for $x$ = 0 and 1, respectively, as shown in Fig. 1(b). Thus, average antiferromagnetic couplings are systematically reduced from the Ga to In compounds in the solid solutions. 

\begin{figure}
\includegraphics[width=8cm]{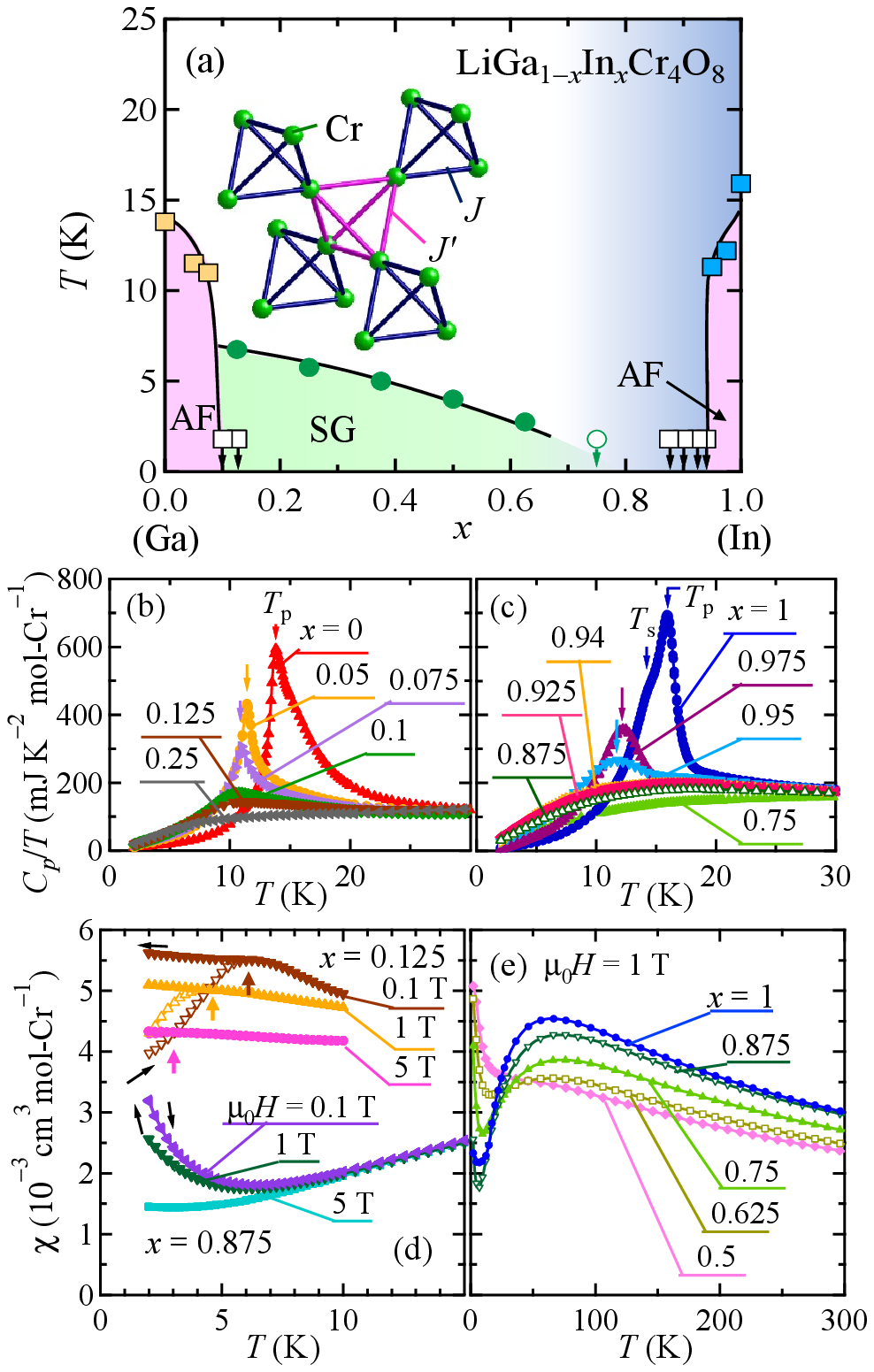}% Here is how to import EPS art
\caption{\label{fig2} 
(color online) (a) $T$$-$$x$ phase diagram of LiGa$_{1-x}$In$_x$Cr$_4$O$_8$ determined by heat capacity $C_p$ and magnetic susceptibility $\chi$ data for polycrystalline samples. `AF' and `SG' denote antiferromagnetic long-range order and spin-glass-like freezing, respectively. The peak temperatures in $C_p$/$T$ are indicated by filled squares, and the spin-glass-like transition temperatures from $\chi$ by filled circles. The open symbols with downward arrows indicate the absence of corresponding anomalies above 2 K. In the shaded region on the In side, spin-gap behavior is observed in $\chi$. The inset shows a BP lattice made up of Cr$^{3+}$ ions. (b, c) Temperature dependences of $C_p$ on the Ga-rich side (b) and on the In-rich side (c). (d) Temperature dependence of $\chi$ for $x$ = 0.125 and 0.875, measured upon heating after zero-field cooling (open marks) and then upon cooling (solid marks) in magnetic fields of 0.1, 1, and 5 T. The spin-glass-like transition temperatures $T_{\mathrm{g}}$ of the $x$ = 0.125 sample are indicated by thick arrows, where the heating and cooling curves separate from each other. (e) Temperature dependence of $\chi$ for $x$ = 0.5, 0.625, 0.75, 0.875, and 1 measured in a magnetic field of 1 T.
}
\end{figure}

The temperature$-$composition phase diagram is shown in Fig. 2(a). Interestingly, the phase transitions of LiGaCr$_4$O$_8$ and LiInCr$_4$O$_8$ are not interconnected smoothly in the solid solutions, in spite of their similar transition temperatures: they are reduced below 2 K by 10 and 6\% substitutions into the Ga and In compounds, respectively. This implies that the ordered states of the end members take substantially different characters. 

Figures 2(b) and 2(c) show the temperature dependences of heat capacity divided by temperature, $C_p$/$T$, for the Ga- and In-rich sides, respectively. The $C_p$/$T$ data of the pure Ga compound exhibit a sharp peak at $T_{\mathrm{p}}$ = 13.8 K, corresponding to a first-order magnetic transition according to recent Li-NMR experiments~\cite{NMR}. Upon increasing the level of In substitution, the peak moves to lower temperatures; $T_{\mathrm{p}}$ = 11.5 and 11 K for $x$ = 0.05 and 0.075, respectively. By further increasing the In content, only a broad peak remains at 10 K for $x$ = 0.1, which eventually disappears for $x$ = 0.125 and 0.25. Thus, In substitution on the Ga-rich side effectively suppresses the long-range ordered state for $x >$ 0.1. 

The $C_p$/$T$ of the In-pure sample shows a sharp peak at $T_{\mathrm{p}}$ = 15.9 K and a shoulder at $T_{\mathrm{s}}$ = 14 K. According to the recent Li-NMR and neutron diffraction experiments, a structural transition from cubic to tetragonal symmetry occurs at $T_{\mathrm{p}}$, followed by long-range magnetic order setting in at $T_{\mathrm{s}}$~\cite{NMR,neutron}. As the Ga content increases, the peak moves to lower temperatures and becomes broader; $T_{\mathrm{p}}$ = 12.2 ($x$ = 0.975) and 11.3 K ($x$ = 0.95). It is then suppressed below 2 K at 0.94. Compared with the Ga-rich side, the ordered state on the In-rich side is thus more rapidly suppressed by substitution. The steep phase boundary at $x$ = 0.94$-$0.95 may indicate the possible presence of a two-phase coexisting region near the boundary composition.

There are a substantial differences in magnetic properties between the Ga- and In-rich sides upon suppression of long-range order: a spin-glass-like phase appears in the wide range 0.1 $\leq x \leq$ 0.625 above 2 K on the Ga-rich side, but is not observed above 2 K on the In-rich side, as shown in Fig. 2(a). Shown for $x$ = 0.125 in Fig. 2(d) are the typical $\chi$'s measured under zero-field-cooled and field-cooled conditions. A large thermal hysteresis is observed in the 0.1 T data below $T_{\mathrm{g}} \sim$ 6 K; the field-cooled data separate from the zero-field-cooled data below $T_{\mathrm{g}}$, and are nearly constant at the maximum value at $T_{\mathrm{g}}$. $T_{\mathrm{g}}$ is reduced with increasing magnetic field; $T_{\mathrm{g}} \sim$ 4.5 and 3 K at $\mu_0$$H$ = 1 and 5 T, respectively. These thermal hystereses and magnetic field dependences are characteristic of a spin-glass transition. However, the presence of a broad peak and the absence of a $T$-linear term in heat capacity distinguish it from conventional canonical spin-glass systems such as dilute magnetic alloys, suggesting that the spin-glass-like transition on the Ga-rich side is due to a spin freezing induced by impurities or lattice defects. Similar transitions have been observed in conventional Cr spinel oxides when impurities are introduced~\cite{Cr-2}.

\begin{figure}
\includegraphics[width=8cm]{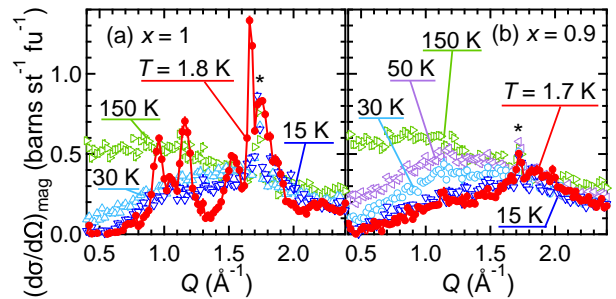}% Here is how to import EPS art
\caption{\label{fig3} 
(color online) Magnetic neutron cross sections for $x$ = 1 (a) and 0.9 (b) measured at various temperatures at the D7 spectrometer at ILL. The peaks marked by asterisks are the 102 magnetic Bragg peak from a small Cr$_2$O$_3$ impurity.}
\end{figure}

Moving on to the In-rich region 0.75 $\leq x \leq$ 0.94, none of the above behaviors is observed above 2 K, as shown for $x$ = 0.875 in Fig. 2(d). Moreover, there is no kink, suggesting the presence of long-range order, but rather a small upturn below $\sim$5 K, which is suppressed with increasing magnetic field. Assuming that it originates from a Curie contribution from orphan spins, the amount of these is estimated to be $\sim$0.2\% of all spins, as will be discussed later.

To obtain a more detailed understanding of magnetic correlations on the In-rich side, polarized neutron scattering experiments were performed. Figure 3 shows the magnetic component of the neutron scattering cross section for the $x$ = 1 and 0.9 polycrystalline samples; the $Q$ dependences of the scattering at 150 K for both samples are form-factor-like, characteristic of an uncorrelated paramagnet. Upon cooling, the scattering at small $Q$ in both samples is reduced, shifting into a broad peak at $Q$ $\sim$ 1.5 \r{A}$^{-1}$ at 15 K, the shape of which appears consistent with the singlet$-$triplet $S$($Q$) for an isolated tetrahedron. From previous inelastic neutron scattering experiments on the $x$ = 1 sample, which show a gapped feature with a similar $Q$-dependence and a gap $\hbar \omega$ = 4.2(1) meV or $\hbar \omega$/$k_{\mathrm{B}}$ = 48(2) K~\cite{neutron}, we expect that the peak observed in the present data is dynamic in origin. 

On further cooling, the spectra of the two samples become entirely different: magnetic Bragg peaks appear at $Q$ = 0.9, 1.2, and 1.7 \r{A}$^{-1}$ in the 1.8 K data for $x$ = 1, indicating the occurrence of long-range magnetic order between 1.8 and 15 K, while the spectrum of $x$ = 0.9 measured at $T$ = 1.7 K remains broad. The upper limit for the magnitude of the ordered moment is as small as 0.05 $\mu_{\mathrm{B}}$/Cr, indicating the absence of long-range order in $x$ = 0.9 above 1.7 K. Neutron diffraction experiments to determine the magnetic structure of the ordered phase in $x$ = 1 are in progress.

\begin{figure}
\includegraphics[width=7.3cm]{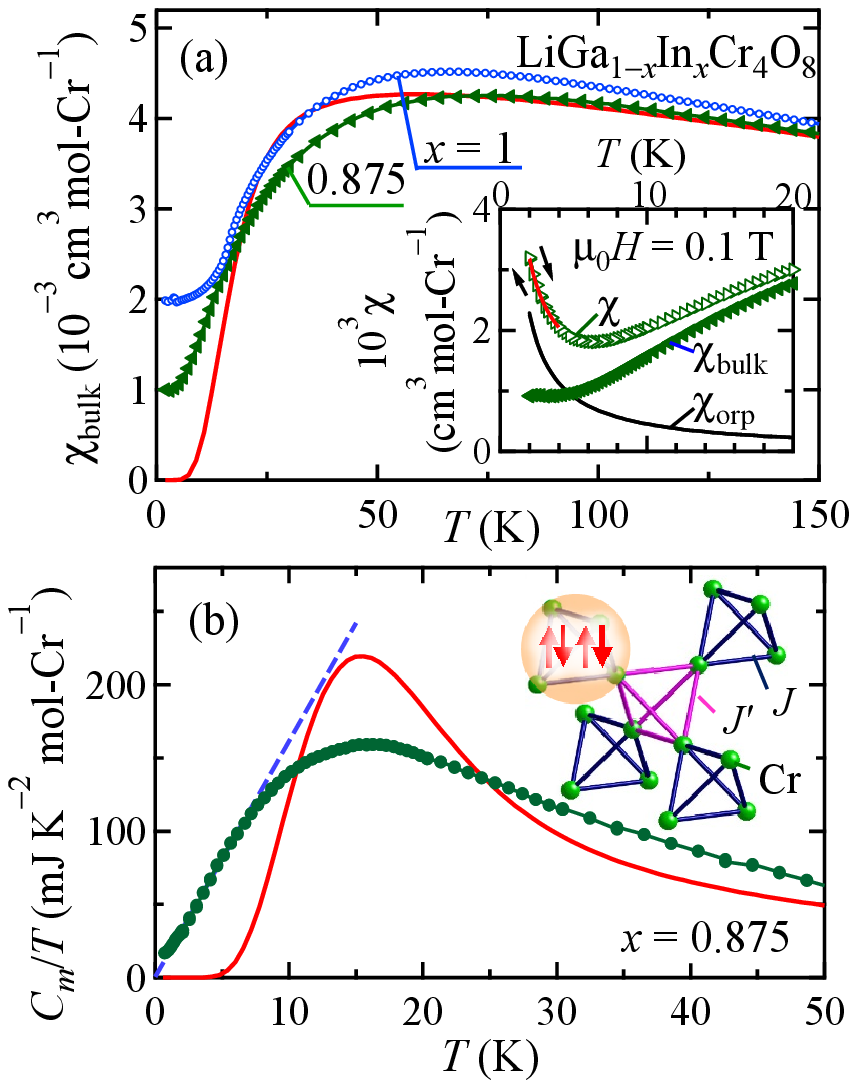}% Here is how to import EPS art
\caption{\label{fig4} 
(color online) Temperature dependence of intrinsic magnetic susceptibility $\chi_{\mathrm{bulk}}$ at $\mu_0$$H$ = 1 T (a) and magnetic heat capacity divided by temperature, $C_m$/$T$, (b) of a polycrystalline sample of LiGa$_{1-x}$In$_x$Cr$_4$O$_8$ with $x$ = 0.875. The data for $x$ = 1 are also shown in (a) for comparison. The solid curves in the main panels of (a) and (b) are the exact results for an isolated $S$ = 3/2 tetrahedron with $J$/$k_{\mathrm{B}}$ = 60 K. The inset in (a) shows the experimental $\chi$ separated into the intrinsic magnetic susceptibility $\chi_{\mathrm{bulk}}$ and $\chi_{\mathrm{orphan}}$ from orphan spins. $C_m$/$T$ in (b) is obtained by subtracting the lattice contribution estimated from the nonmagnetic isomorphic compound LiInRh$_4$O$_8$ from the total $C_p$/$T$. The broken line shows a linear fit of the 1$-$5 K data. The inset in (b) schematically depicts a tetramer singlet state realized on the small tetrahedron in the BP lattice.}
\end{figure}

Combining all the experimental data mentioned above, LiGa$_{1-x}$In$_x$Cr$_4$O$_8$ with $x \sim$ 0.9 shows neither long-range magnetic order nor spin-glass-like freezing down to 2 K. Hereafter, we focus on the magnetic properties of $x \sim$ 0.9. Figure 4(a) shows the temperature dependence of the intrinsic magnetic susceptibility $\chi_{\mathrm{bulk}}$ of the $x$ = 0.875 sample with $\theta_{\mathrm{W}}$ = $-$342 K and $g$ = 2.01, which has been obtained by subtracting the orphan spin contribution $\chi_{\mathrm{orphan}}$ from $\chi$. $\chi_{\mathrm{orphan}}$ was estimated by fitting the $\chi$ data between 2 and 4 K to the equation $\chi$ = $\chi_{\mathrm{bulk}} + \chi_{\mathrm{orphan}}$ = $\chi_{\mathrm{const}} + C_{\mathrm{orphan}}$/($T - \theta_{\mathrm{orphan}}$). 
The fitting result is satisfactory, as shown in the inset of Fig. 4(a) for $C_{\mathrm{orphan}}$ = 4.5(2) $\times$ 10$^{-3}$ cm$^{3}$ K mol$-$Cr$^{-1}$, $\theta_{\mathrm{orphan}}$ = $-$0.01(5) K, and $\chi_{\mathrm{const}}$ = 9.1(3) $\times$ 10$^{-4}$ cm$^3$ mol$-$Cr$^{-1}$. Assuming $g$ = 2.01, this means that only 0.2\% of all $S$ = 3/2 spins behave as nearly free orphan spins, which may originate from defects in the crystal. 

The $\chi_{\mathrm{bulk}}$ values of $x$ = 0.875 and 1 rapidly decrease upon cooling below $\sim$70 K owing to the presence of a gap or a pseudogap in the spin excitation spectrum. This gaplike decrease in $\chi$ becomes weaker with further increasing Ga content, as shown in Fig. 2(e); the $\chi$ of $x$ = 0.5 monotonically increases with decreasing temperature. The size of the gap in $x$ = 0.875 is estimated to be $\Delta$/$k_{\mathrm{B}}$ = 60 K by fitting the $\chi_{\mathrm{bulk}}$ data between 4 and 350 K to the exact result for a spin-3/2 isolated tetrahedron, similar to $\Delta$/$k_{\mathrm{B}}$ = 57 K for $x$ = 1~\cite{BP-2}. At low temperatures, the decrease in the $x$ = 1 data weakens below the ordering temperature at $\sim$14 K, while the $\chi_{\mathrm{bulk}}$ of $x$ = 0.875 continues to decrease without any sign of magnetic order, approaching a small value of $\sim$10$^{-3}$ cm$^{3}$ mol$-$Cr$^{-1}$ at 2 K. Thus, the spin gap of $x$ = 0.875 may be filled with a certain amount of in-gap states, although the residual $\chi_{\mathrm{bulk}}$ value is approximate because of the ambiguity in the subtraction of the orphan spin contribution. 

The magnetic heat capacity divided by temperature, $C_m$/$T$, of the $x$ = 0.875 sample exhibits a broad peak at $T \sim$ 16 K, as shown in Fig. 4(b). The calculated curve for an $S$ = 3/2 isolated tetrahedron with $\Delta$/$k_{\mathrm{B}}$ = $J$/$k_{\mathrm{B}}$ = 60 K shows a maximum at 15 K, which coincides with the maximum at 16 K of the experimental data. This suggests that the associated entropy release is due to the opening of a spin gap with the order of $J$. Compared with the calculated curve, however, the experimental curve is much broader. The $C_m$/$T$ does not decrease exponentially but is proportional to $T$, i.e., $C_m \propto T^2$, at 1$-$5 K. A linear fit of the 1$-$5 K data yields $C_m$/$T$ = 0.7(7) + 16.1(2)$T$. The negligibly small first term indicates that there is little $T$-linear $C_m$ such as that observed in metals or spin glass systems. 

The quantum spin-1/2 Heisenberg antiferromagnet on the uniform pyrochlore lattice is theoretically expected to have a finite spin gap instead of a long-range magnetic order, while the nature of its ground state is still subject to debate~\cite{theo-1,theo-2,theo-3}. The BP lattice has been studied in the strong coupling approximation, originally with a view to understanding the ground state in the uniform one~\cite{theo-1,theo-2}. In the case of $J^{\prime}$ = 0, it is clear that a twofold tetramer-singlet ground state with a gap $\Delta$ = $J$ to the first excited triplet is realized~\cite{tetramer}. This tetramer singlet can be broken into two dimer singlets by switching on $J^{\prime}$. In a mean-field approximation, the ground state at $J$ = $J^{\prime}$ is expected to be a valence bond crystal of dimers, in which three of four tetrahedron sublattices have an ordered dimer-pair pattern with the rest remaining disordered~\cite{theo-1}, giving the in-gap nonmagnetic states~\cite{theo-2}. The origin of the quadratic $C_m$ in the $x \sim$ 0.9 sample may be understood by the presence of a similar partially disordered state.

However, the observed finite $\chi_{\mathrm{bulk}}$ at 2 K in the $x \sim$ 0.9 samples indicates that there are a significant number of magnetic states in addition to nonmagnetic states in the spin gap, which seems to contradict the above theoretical expectation. The question is whether these magnetic states are intrinsic to the BP or induced by disorder. The fact that the $x$ = 0.875 sample does not show a spin-glass freezing above 2 K and has rather few orphan spins of $\sim$0.2\% suggests that disorder effects in the $x \sim$ 0.9 samples are minor compared with those on the Ga-rich side. Nevertheless, one has to be careful to exclude the possibility of disorder effects in frustrated antiferromagnets because of the close proximity of many competing states to the ground state. There is also a possibility that the two-phase coexistence complicates the interpretation. If the magnetic in-gap states are intrinsic to the BP antiferromagnet, they must be ascribed to nonlocal magnetic excitations caused by correlations between tetramer singlets. Future NMR and neutron experiments should uncover the interesting properties of the BP antiferromagnets.

In summary, we find that neither a long-range order nor a spin-glass-like freezing is observed down to 2 K at $x \sim$ 0.9 in LiGa$_{1-x}$In$_x$Cr$_4$O$_8$. Instead, a ``pseudo'' spin gap behavior is observed, as evidenced by a finite $\chi_{\mathrm{bulk}}$ at 2 K and a quadratic $C_m$ at 1$-$5 K. This unusual state may be related to weakly coupled tetramer singlets on a BP lattice with geometrical frustration and bond alternation. 

%\section*{Acknowledgments}

We thank Y. Tanaka, M. Yoshida, M. Takigawa, and Y. Motome for helpful discussion. This work was partly supported by JSPS KAKENHI Grant Number 25800188.

\end{document}